# Orbital Magnetization Measurement of the Quantum Hall to Insulator Transition

D.R. Faulhaber, and H.W. Jiang

Department of Physics and Astronomy,

University of California, Los Angeles, California 90095

## Abstract

We present a magnetization measurement probing the transition from a quantum Hall to insulating (QH-I) state for a two-dimensional electron gas in a disordered GaAs/AlGaAs heterostructure. Using a highly sensitive DC torque magnetometer, we discover an abrupt change in the orbital magnetization precisely at the critical point for the QH-I transition. Since this transition is predicted to be a second order quantum phase transition, a thermodynamic signature in magnetization is totally unexpected. The observed feature is reminiscent of the well-known de Haas-van Alphen oscillations arising from discontinuous jumps of the chemical potential as Landau Levels are successively populated.

In disordered quantum Hall systems, phase transitions from a quantum Hall to insulating state are observed when either the magnetic field or electron density is varied.[1] At zero temperature, a QH state, characterized by a zero longitudinal resistivity $\rho_{xx}$, undergoes a transition to insulator ($\rho_{xx} = \infty$), at a given critical field or density. Transitions to the insulating state can occur from a number of different QH states, and are accepted as being examples of continuous quantum phase transitions.[2] An enormous amount of work studying the various QH-I transitions, primarily in the form of transport, has yielded much information in this regime[1,3]; however, nothing has been done to explore the thermodynamic ground state properties.

In this work, we measure the orbital magnetization of a 2D electron gas (2DEG), focusing around the QH-I transition in a GaAs/AlGaAs heterostructure. For a given system of $N$ particles, the magnetization is defined as: $M = -\left(\frac{\partial F}{\partial B}\right)_N$, where $F$ is free energy. Since $F$ reflects the many-body interaction effects of the 2D system, magnetization can be regarded as an interaction-renormalized thermodynamic quantity probing the ground state properties of a system in response to an external magnetic field. Since the QH-I is a second order quantum phase transition, the free energy and its derivative are expected to vary smoothly across the critical point, leaving no signature in thermodynamic quantities, like magnetization. In this letter, we report the surprising experimental observation of a finite temperature magnetization signature at the QH-I transition.

The magnetometer is modeled after a design by Wiegers[4], with modifications for simultaneous transport measurement. As shown in Fig. 1, an epoxy rotor suspended by a .001'' diameter phosphor bronze wire is positioned at the center of an epoxy stator. The rotor has a set of Ag electrodes evaporated on its outside perimeter while the inside perimeter of the stator has two sets. The sample is mounted on a stage and placed on a platform that bisects the rotor. In this configuration the normal to the sample lies at an angle of $\approx 40$ degrees with respect to the field axis. A small copper coil wrapped around the stage provides a known magnetic moment for calibration. In addition, two other .0005'' copper leads used to make contact to the sample are fixed to the stage and carefully twisted along the axis of the rotor. Since the electrons are constrained to two dimensions, their motion in a magnetic field will yield an orbital magnetic moment perpendicular to the surface of the sample. This moment produces a torque given by, $\vec{\tau} = (\vec{M} \times \vec{B})$ and the rotation resulting from this torque is detected by a standard AC bridge method. The two capacitances between rotor and stator along with a ratio transformer complete the bridge circuit. Any change in capacitance between the electrodes, corresponding to a change in the overlap of surface area between them as the wheel rotates, is detected at the lock-in amplifier.[4] The magnetometer is mounted on a brass stage that is centered in a superconducting magnet capable of fields up to 10T. The maximum torque sensitivity obtained is a few times $10^{-14}$ N-m at a magnetic field of 1 T.

The samples used are GaAs/Al$_{0.3}$Ga$_{0.7}$As heterostructures grown by molecular-beam eptaxy. We obtain sufficient disorder making the QH-I transition accessible from a $1.5x10^{12} cm^{-2}$, δ-doped Si layer at a setback distance of 70 Å to the 2DEG. An electrical contact to the 2DEG is made by diffusing indium on a corner of the sample and an evaporated aluminum gate covers the remaining $4x4mm^2$ surface. With the gate we apply a voltage and tune the density of carriers of the 2DEG. The mobility of our sample is measured to be 186,000 $cm^2/V \cdot s$, with a zero gate voltage density of $6x10^{11} cm^{-2}$.

In Fig. 2, we present plots of the magnetization as a function of density at 1.5K for several magnetic fields. Measurements are made by fixing the applied magnetic field perpendicular to the sample and sweeping the density. The major advantage of sweeping the density instead of magnetic field for this experiment is the quenching of the magnetic moments from the device and substrate after thermal and magnetic equilibrium are reached. These effects often swamp the small signal of the electrons while sweeping the magnetic field and much effort must be given to subtract them[4,5,6]. Standard AC magneto-capacitance between the gate electrode and the 2DEG is taken simultaneously, allowing us to obtain the density of states as well as the transport of the 2DEG.[7]

Saw-tooth like oscillations are observed and consistent with previous experiments.[5,6,8,9] These oscillations are the well-known de Haas-van Alphen (dHvA) oscillations corresponding to discontinuous jumps of chemical potential as the Landau Levels are successively populated. The positions of these magnetization signatures identified as positions at even integer filling factors match well with the simultaneous transport data. Oscillations of the magnetization are clearly resolved and their magnitude decreases while reducing the density as expected from a smaller number of particles present. The disorder broadening of the landau levels is apparent in the data where the predicted saw-tooth form is smeared out as expected for a lower mobility sample. The broadening tends to increase at a lower density, which could be a consequence of a decrease in carrier mobility or increase in disorder. The amplitude of the oscillations at even filling factors is near the expected $2\mu_B^* N$, where $\mu_B^* = \dfrac{e\hbar}{2m^*}$ is the effective Bohr magneton, $m^* = .0667 m_e$ for GaAs. This value corresponds roughly to $1.33 x 10^{-11} J/T$ for a density of $3 x 10^{11} cm^{-2}$. At higher fields, in Figs. 2c, 2d, the spin splitting can be resolved at ν =3, 5.

In addition to the magnetization signatures attributed to the dHvA oscillations occurring precisely between QH states, we observe an additional feature. This new feature is circled in Fig 2 in the low-density regime for ν < 2. Unlike the dHvA oscillations discussed above, this occurrence is completely unexpected. The possibility of this feature being attributed to an oscillation occurring at ν = 1 can be ruled out. In this rather disordered sample, the Landau levels are spin-degenerate at low densities. In other words, the small spin split at ν = 1 cannot be resolved[10], and simultaneous transport measurement confirms this. Furthermore, we estimate the amplitude of the new magnetization jump to be roughly $\mu_B^* N$ which is far greater than any spin splitting[6,8].

In Fig. 3, we plot the positions of the unexpected feature in the density-magnetic field plane along with the QH phase boundaries (extended states) obtained from transport measurements. Following the convention[11], the QH and the insulating states are specified by their Hall conductance, $S_{xy} = \sigma_{xy} h/e^2$ with $S_{xy}$ an integer. The new magnetization features are represented by red triangles, and we find that they directly correspond to the phase boundary to insulating state, $S_{xy} = 2 \rightarrow S_{xy} = 0$ (2-0) transition. This result is surprising. We do expect to see magnetization signatures precisely at the center of a QH phase at even integer filling factors as indicated by the dHvA oscillations in the figure. However, the feature occurring at the 2-0 transition happens precisely at the lowest extended state, and has no explanation. In fact for all other phase boundaries in Fig 3 (i.e., 2-4, 4-6 etc) there is no evidence of a magnetization signature. Since the new feature is an unexpected observation, we choose to assign its location close to the peak, similar to measured positions of dHvA oscillations. Error bars arise from averaging a large number of scans for fixed field values. Changing the method of defining the position of the signature simply shifts the data up or down a negligible amount within the error bars. It is apparent that within an acceptable error no matter where we choose the location of the feature, it is associated with the 2-0 transition.

We have also investigated the temperature evolution of this new feature at a fixed field of 3.5 T. As we increase the temperature, thermal broadening begins to overwhelm the small signal, which is completely indistinguishable at about 10 K. If this new signature can be associated with a gap in the energy spectrum like dHvA effect, we infer a gap size on the order 1/7 the cyclotron energy.

To illustrate why the observed feature is unexpected, we plot the expected magnetization for a non-interacting 2DEG in Fig. 4b, along with our experimental observation in Fig 4a. The curve is calculated numerically by using a Gaussian broadened density of states of the following form:

$$DOS(\varepsilon, B) = \frac{eB}{\pi\hbar} \frac{1}{\sqrt{2\pi}\Gamma} \sum_n \exp(-\frac{(\varepsilon-\varepsilon_n)^2}{2\Gamma^2}),\qquad(2)$$

where $\Gamma$ is the rms half-width and is taken to be proportional to $\sqrt{B}$, and $\varepsilon_n = (n+\frac{1}{2})\frac{e\hbar B}{m^*}$ is the Landau level energy. The theoretical curve clearly captures the essence of the experimental data at high density for filling factors $\nu$ = 2, 4, and 6. In contrast, the feature at low-density is not expected in this non-interacting particle model. The magnetization should tend monotonically to zero as the density approaches zero, following the drop in magnetization at $\nu$ = 2.

The QH-I transition is commonly thought of as being as zero temperature, second order quantum phase transition. The fluctuation of the states at finite temperature should only be manifested in transport experiments.[2]. The fact that the no change in the free energy is expected implies a quantity like magnetization should be continuous. Since the observation we have made is so similar to the dHvA oscillations; it suggests there is a gap in the energy spectrum. Our observation leads us to speculate a first order phase transition is occurring; however, we offer no theoretical explanation for this discrepancy.

The finding here also challenges the notion of the QHE-I transition being a special example of the transitions between different QH states (i.e., the plateau - plateau transitions). In the theoretical framework for the global phase diagram of the quantum Hall effect[11], all transitions between QH states and QH-I states are in the same class. In Fig. 3 we see the plateau - plateau transitions between two QH states yield no thermodynamic signatures in magnetization, whereas the QH-I leaves a definite and unambiguous thermodynamic signature. It is possible the spin-degenerate nature of the 2-0 transition is a special type of QH-I transition, [12,13] we intend to study other QH-I transitions (i.e. 1-0, 1/3 -0 etc.) to fully map out the observed feature in the global phase diagram.

Finally, we would like to comment on the possible relevance of the observed effect to the QH percolation picture.[14] In a QH state we find two contributions to the total magnetization, arising from the quantization of orbital motion for individual Landau levels (bulk currents), and currents induced from the existence of a large potential at the sample boundary (edge currents). Both bulk and edge currents circulate around the sample boundary, with the sum of the two yielding the total magnetic moment contribution.[15] In the context of the percolation picture the large bulk currents break when transition to insulating phase occurs, splitting into a number of smaller loops spread throughout the 2D area. The magnetization contribution from the sum of these individual currents should match the initial. We are witnessing an effect contrary to this, precisely at the transition the magnetization is discontinuous. This implies the ground state of the system may be different form what the percolation picture predicts.

We summarize by stating that a direct orbital magnetization measurement of the 2-0 transition in a quantum Hall system has yielded a definite and unexpected signature. The signature manifests itself as an additional oscillation in magnetization precisely at the critical point to insulating state. This finding is directly at odds with existing theories, and no theoretical explanation as of yet has been proposed for its existence.

We would like to thank S. Kivelson, R. Narasimhan and L. Pryadko for discussions and B. Alavi for technical assistance. This work is supported by NSF under grant \# DMR 0071969.

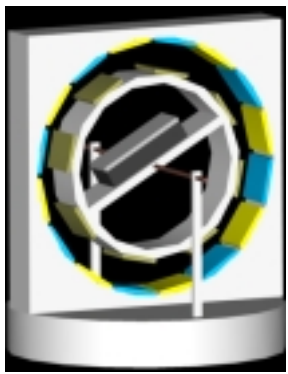

Figure 1: Schematic of the epoxy magnetometer modeled after original design by Wiegers et al, with gated sample mounted on the platform bisecting the wheel. Sweeping the gate voltage at fixed magnetic fields changes the density of the 2DEG, which leads to a change in orbital magnetization. The induced torque rotates the wheel and leads to a change in overlap of surface area between electrodes on the wheel and those on the house. This change is detected by an AC bridge technique.

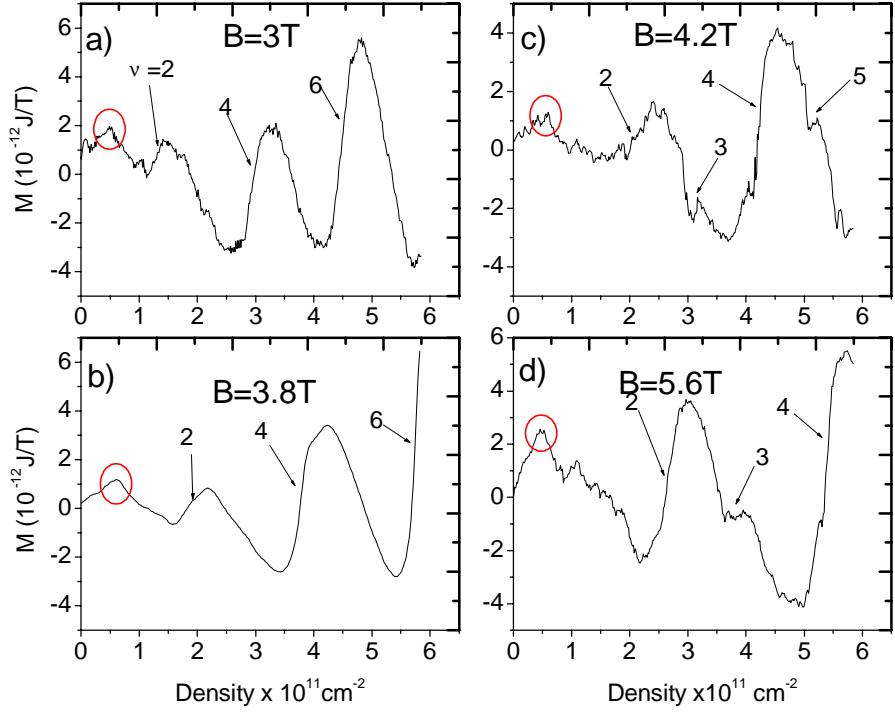

Figure 2: Plots of magnetization oscillations as a function of density for four different fields at 1.5 K, with positions of filling factors indicated. The magnitude of oscillations near even integer filling factors is close to their expected value of $2\mu_B^* N$. The surprise comes at low density as after the ν = 2 oscillation, the data goes through an additional oscillation (circled), and then tends to zero.

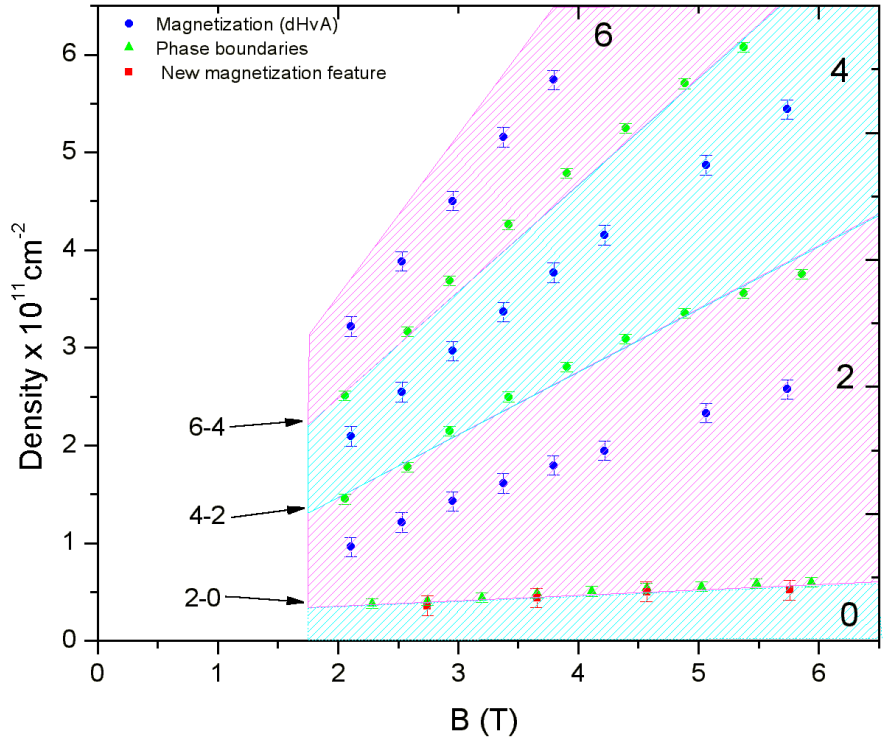

Figure 3: Phase diagram for 2DEG in the density-field plane. Shaded regions correspond to QH states, $S_{xy}$ = 0, 2,4,6 where boundaries (triangles) are taken from transport. The dHvA oscillations (circles) occur at the center of each QH state. At the lowest phase boundary (2-0 transition) we do not expect a signature in magnetization however, one is apparent (squares).

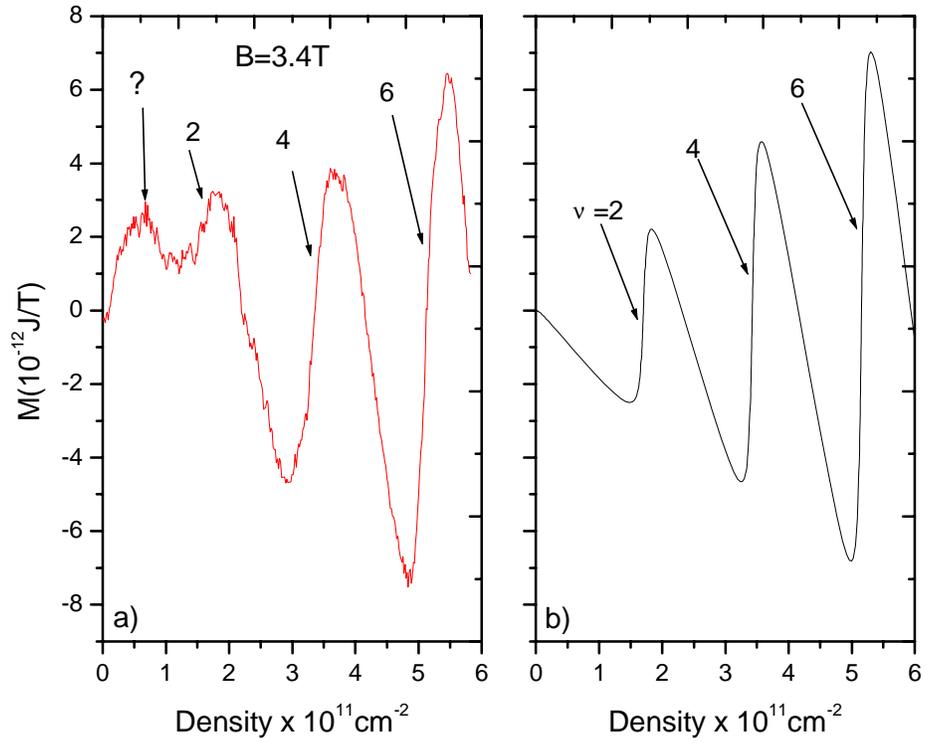

Figure 4: a) Experimental magnetization data taken at B= 3.4T. b) Numerical calculation of magnetization as a function of density for the non-interacting particle model. Comparing the two figures, one can see that the observation indicated by the question mark is not expected.